# CT sinogram-consistency learning for metal-induced beam hardening correction

Hyoung Suk Park

*Division of Integrated Mathematics,*

*National Institute for Mathematical Sciences, Daejeon, 34047, Korea.*

Sung Min Lee, Hwa Pyung Kim, Jin Keun Seo[*]

*Department of Computational Science and Engineering,*

*Yonsei University, Seoul, 120-749, South Korea*

Yong Eun Chung

*Department of Radiology, Yonsei University College of Medicine, Seoul, 03722, South Korea*

## Abstract

**Purpose:** This paper proposes a sinogram-consistency learning method to deal with beam-hardening related artifacts in polychromatic computerized tomography (CT). The presence of highly attenuating materials in the scan field causes an inconsistent sinogram, that does not match the range space of the Radon transform. When the mismatched data are entered into the range space during CT reconstruction, streaking and shading artifacts are generated owing to the inherent nature of the inverse Radon transform

**Methods:** The proposed learning method aims to repair inconsistent sinogram by removing the primary metal-induced beam-hardening factors along the metal trace in the sinogram. Taking account of the fundamental difficulty in obtaining sufficient training data in a medical environment, the learning method is designed to use simulated training data and a patient-type specific learning model is used to simplify the learning process.

**Results:** The feasibility of the proposed method is investigated using a dataset, consisting of real CT scan of pelvises containing simulated hip prostheses. The anatomical areas in training and test data are different, in order to demonstrate that the proposed method extracts the beam hardening features, selectively. The results show that our method successfully corrects sinogram inconsistency by extracting beam-hardening sources by means of deep learning.

**Conclusion:** This paper proposed a deep learning method of sinogram correction for beam hardening reduction in CT for the first time. Conventional methods for beam hardening reduction are based on regularizations, and have the fundamental drawback of being not easily able to use manifold CT images, while a deep learning approach has the potential to do so.

## I. INTRODUCTION

In computerized tomography (CT), the presence of highly attenuating materials such as metal, concentrated iodinated contrast media or bone complicates reconstruction[1] by violating an assumption of the forward model: that sinogram data are the Radon transform of an image. The increasing use of metallic implants in a generally aging population makes metal-induced artifacts a major impediment to CT diagnosis. The mismatched projection data due to these effects cause severe streaking and shading artifacts in the reconstructed CT images, as shown in Fig. 1. Metal artifacts are caused by the effects of beam-hardening of polychromatic X-ray photon beams and the various complicated metal-tissue interactions such as scattering, nonlinear partial volume effects and electric noise. Although extensive research efforts have sought to improve CT reconstruction methods, tackling metal-related artifacts is a very challenging problem because the inconsistent data induced by metal depend non-linearly on the geometries and placements of the metal objects.

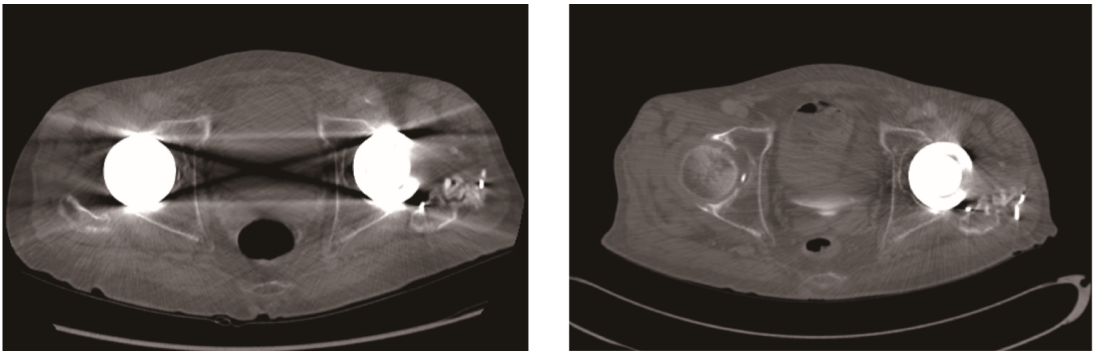

FIG. 1. CT images for patients in the presence of high-attenuating materials (hip prosthesis).

This paper uses a deep learning approach to repair sinogram inconsistencies, so as to minimize artifacts induced by highly attenuating material. The proposed learning method aims to correct only the main sources of beam-hardening artifacts, while leaving fine data structures intact. We corrects the primary beam-hardening factor of metals along a sinusoidal metal trace. To simplify the learning process, we consider a patient-type specific learning model (e.g., learning beam hardening artifacts caused by hip prosthesis). This paper explores the feasibility of the proposed method using simulated training dataset, due to difficulty in collecting training data from patients (i.e., pairs of artifact-free and artifact-contaminated CT images). The anatomical areas in training and test data are different, in order to demonstrate that the proposed method extracts the beam hardening features, selectively; for example, the test data comes from sinograms corresponding to CT images of the pelvis, whereas the training sinograms are obtained from the CT images of abdomen and chest. (See Fig. 5 in the results section.) The beam-hardening factors due to metals are expected to lie on a very-low-dimensional manifold, thus allowing extraction of beam-hardening features in a supervised way. We also apply the proposed method to the projection data acquired in different acquisition environments from those used for training; for example, data acquired with different energy spectrum or data that includes (approximated) scatter. We employ a U-net[20] for learning the nonlinear beam-hardening features.

To be precise, let $\boldsymbol{x}(\varphi, s)$ (input data for deep learning) represent the projection data at projection angle $\varphi \in [0, 2\pi)$ and detector position $s \in \mathbb{R}$. For simplicity, we shall restrict to considering two-dimensional parallel-beam CT. According to the Lambert-Beer's law[2,10], the CT sinogram data $\boldsymbol{x}$ can be expressed as

$$\boldsymbol{x}(\varphi, s) = -\ln\left(\int \eta(E) \exp\left\{-[\mathcal{R}(\mu_E)](\varphi, s)\right\} dE\right), \qquad (1)$$

where $\mu_E$ denotes the attenuation coefficient distribution at photon energy level $E$, $\eta(E)$ represents fractional energy at $E$[8,18], and $\mathcal{R}(\mu_E)$ is the Radon transform of $\mu_E$. See Fig. 2. Let $Y$ denote the range space of the Radon transform $\mathcal{R}$. Then, $Y$ can be viewed as a subspace of the sinogram space $X$.

The goal is to learn the sinogram correction function $f : X \to Y$ such that $\mathcal{B}_{\text{FBP}}(f(\boldsymbol{x}))$ is artifact-removed image, where $\mathcal{B}_{\text{FBP}}$ represents the filtered backprojection (FBP) operator[4]. Noting that the orthogonal projection of $\boldsymbol{x}$ onto $Y$ is $\mathcal{P}\boldsymbol{x} = \mathcal{R}(\mathcal{B}_{\text{FBP}}(\boldsymbol{x}))$, we have $\mathcal{B}_{\text{FBP}}(\boldsymbol{x}) = \mathcal{B}_{\text{FBP}}(\mathcal{P}\boldsymbol{x})$ and therefore $f = \mathcal{R}\mathcal{B}_{\text{FBP}}$ is not appropriate for MAR.

The proposed method uses a U-Net to learn the sinogram-correction function $f : \boldsymbol{x} \mapsto \boldsymbol{y}$ from a training data $\{(\boldsymbol{x}^{(i)}, \boldsymbol{y}^{(i)}) : i = 1, \cdots, N\}$ where each $\boldsymbol{x}^{(i)}$ is an inconsistent sinogram in the presence of a highly attenuating material and $\boldsymbol{y}^{(i)}$ is the corresponding sinogram in the absence of the highly attenuating material. Roughly speaking, the $f$ can be achieved from

$$f = \operatorname*{argmin}_{f \in \mathbb{U}_{net}} \frac{1}{N} \sum_{i=1}^{N} \|f(\boldsymbol{x}^{(i)}) - \boldsymbol{y}^{(i)}\|^2 \qquad (2)$$

where $\mathbb{U}_{net}$ is a deep convolutional neural network with some domain knowledge.

Various numerical experiments show that the proposed deep learning method provides the proper correction of the inconsistent sinogram with preserving its fine-structure. To confirm the goodness of the sinogram repair, we use the above-mentioned method for evaluation of the output $f(\boldsymbol{x})$. The results suggest that the proposed deep learning method using simulated training data can appropriately correct an inconsistent sinogram for MAR.

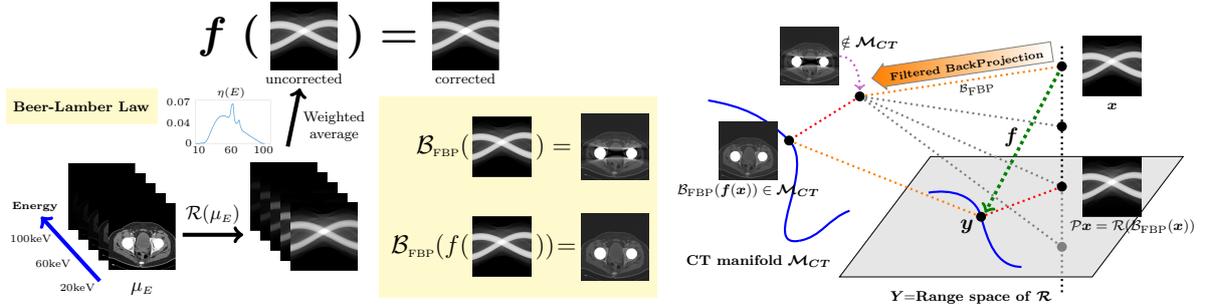

FIG. 2. Schematic diagram of data inconsistency correction in polychromatic X-ray CT. (Left bottom) Images of $\mu_E$ over 20keV$\leq E \leq$ 150 keV and the corresponding Randon transforms. (Left) The sinogram data is approximately a weighted sum of the Radon transforms. Data inconsistencies are caused by the energy-dependent behavior of the attenuation distribution. (Middle-bottom) Filtered backprojections $\mathcal{B}_{\text{FBP}}$ of the uncorrected $\boldsymbol{x}$ and corrected data $f(\boldsymbol{x})$. (Right) Inconsistency artifact is related to the degree of discrepancy between $\boldsymbol{x}$ and the range space $Y$ of Radon transform. We need to extract the source of artifacts in the data $\boldsymbol{x}$ by learning the features considering CT image information (i.e. CT manifold $\mathcal{M}_{\text{CT}}$).

## II. METHOD

### A. Learning objectives and U-net

Let $\boldsymbol{x} \in X$ represent a sinogram data in two-dimensional parallel-beam CT, where $X = L^2([0, 2\pi) \times \mathbb{R})$ is the space of square integrable functions with its norm $\|\boldsymbol{x}\| = \sqrt{\int_0^{2\pi} \int_{\mathbb{R}} |\boldsymbol{x}(\varphi, s)|^2 \, ds d\varphi}$. Let $Y$ be the range space of Radon transform $\mathcal{R}$, a subspace of $X$. To avoid notational complexity and for ease of explanation, we use the same notations $(\boldsymbol{x}, \boldsymbol{y}, X, Y)$ to represent their discrete versions, used in the practice.

In the presence of the high attenuation materials such as metallic objects, $\boldsymbol{x}$ may not fit with $Y$ and the inconsistency of $\boldsymbol{x}$ can be quantified by

$$d_{\boldsymbol{x}} := \min_{\mu \in L^2(\mathbb{R}^2)} \|\mathcal{R}(\mu) - \boldsymbol{x}\|. \quad (3)$$

The mismatch $d_{\boldsymbol{x}}$ is somehow related to the degree of artifacts in the image reconstructed by the FBP method. Denoting FBP operator by $\mathcal{B}_{\text{FBP}}$, the reconstructed image $\mathcal{B}_{\text{FBP}}(\boldsymbol{x})$ can be viewed as the least-square fitting solution, in the sense that

$$\mathcal{B}_{\text{FBP}}(\boldsymbol{x}) := \underset{\mu \in L^2(\mathbb{R}^2)}{\operatorname{argmin}} \|\mathcal{R}(\mu) - \boldsymbol{x}\| \quad (4)$$

where 'argmin' stands for the argument of the minimum. Hence, the projection operator from $X$ onto $Y$ is expressed as $\mathcal{P} = \mathcal{R}\mathcal{B}_{\text{FBP}}$.

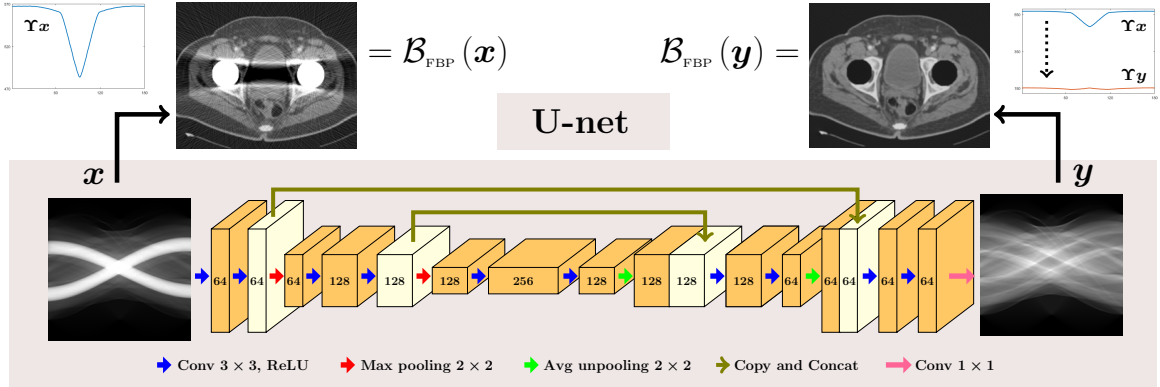

FIG. 3. Deep learning architecture relying on operations of convolutional and pooling layers, and rectified linear uints (ReLU). To address the difficulties of collecting training data in a CT clinical environment, we use a patient-type specific learning model (e.g. hip replacements) and simplify the learning process.

The goal is to find a suitable correction map $f : \boldsymbol{x} \mapsto \boldsymbol{y}$ for metal artifact reduction in such a way that $\boldsymbol{y} \approx \mathcal{R}(\mu_{E_*})$ with $\mu_{E_*}$ being the attenuation distribution at a fixed energy $E_*$. Imagine that $\mathcal{M}_{\text{CT}}$ is a manifold near which various monochromatic CT images are residing. Here, it is assumed that the data of monochromatic CT image-generating distribution is concentrated near $\mathcal{M}_{\text{CT}}$ embedded in the space $\mathbb{R}^{256 \times 256}$. The dimensionality of $\mathcal{M}_{\text{CT}}$ can be much lower than $256 \times 256$, because the probability of generating a CT-like image by choosing pixel intensities independently of each other is close to zero[3]. This manifold $\mathcal{M}_{\text{CT}}$ should not include images having serious metal artifacts, as shown in Fig. 2.

The map $f$ must take account of this CT manifold $\mathcal{M}_{\text{CT}}$, because $\mathcal{M}_{\text{CT}}$ is regarded as prior information of CT images. Hence, $f(\boldsymbol{x})$ should be determined by not only $\boldsymbol{x}$ but also the manifold $\mathcal{M}_{\text{CT}}$. Due to highly nonlinear and complicated structure of $\mathcal{M}_{\text{CT}}$, it could be very difficult to find $f$ without using machine learning techniques.

We adopt U-net to learn $f$. Denoting by $f_{\text{eed}}$ the feed forward map corresponding to $f$, the relation between an input vector $\boldsymbol{x}$ and an output vector is expressed as

$$\boldsymbol{y} = f(\boldsymbol{x}) = f_{\text{eed}}(\boldsymbol{x}; \boldsymbol{w}_1, \cdots, \boldsymbol{w}_L) \tag{5}$$

where $\boldsymbol{w}_l$ is the vector including the weight and bias for $l^{th}$ layer. We collect a set of training data $\{(\boldsymbol{x}^{(i)}, \boldsymbol{y}^{(i)})\}_{i=1}^N$ and determine the vectors $\boldsymbol{w}_1, \cdots, \boldsymbol{w}_L$ by solving

$$\underset{(\boldsymbol{w}_1, \cdots, \boldsymbol{w}_L)}{\text{argmin}} \frac{1}{N} \sum_{i=1}^N \|f_{\text{eed}}(\boldsymbol{x}^{(i)}; \boldsymbol{w}_1, \cdots, \boldsymbol{w}_L) - \boldsymbol{y}^{(i)}\|^2 \tag{6}$$

The most fundamental challenge is to collect enough medical image data to be well-labeled $\{(\boldsymbol{x}^{(i)}, \boldsymbol{y}^{(i)})\}$. This problem seems to be very difficult to solve. To address this fundamental disadvantages, we develop a patient-type specific learning method, which simplify the way for learning $f$ from a simulated data. The proposed method fixes only major beam-hardening sources along the metal trace and does not touch relatively minor artifact sources. Noting that the nonlinear beam-hardening factor is determined mainly by the geometry and the arrangement of the metal objects, it is possible to learn the sinorgam correction from simulated data. Fig. 3 depicts a rough schematic of this approach.

The network consists of a contracting path and an expansive path. The contracting path contains convolutions with a size of $3 \times 3$, each followed by a rectified linear unit (ReLU), along with $2 \times 2$ max pooling for downsampling. The expansive path used $2 \times 2$ average unpooling[22] instead of max pooling for upsampling. It was concatenated with the correspondingly cropped feature from the contracting path. At the last layer a $1 \times 1$ convolution is used to reduce the dimensions.

## B. Sinogram inconsistency effects

The data consistency condition for sinogram $\boldsymbol{x}$ involves the following moment condition: for any integer $k \geq 0$, the function $\int_{\mathbb{R}} s^k \boldsymbol{x}(\varphi, s) ds$ with respect to $(\cos\varphi, \sin\varphi)$ is a homogeneous polynomial of order $k$[14].

For simplicity, we only consider the zeroth order moment consistency condition. Define

$$\Upsilon\boldsymbol{x}(\varphi) := \int_{\mathbb{R}} \boldsymbol{x}(\varphi, s) ds \qquad (7)$$

This $\Upsilon\boldsymbol{x}$ can be divided into its low frequency part $\Upsilon\boldsymbol{x}_L$ and its high frequency part $\Upsilon\boldsymbol{x}_H$:

$$\Upsilon\boldsymbol{x}(\varphi) = \Upsilon\boldsymbol{x}_L(\varphi) + \Upsilon\boldsymbol{x}_H(\varphi), \quad \varphi \in (0, 2\pi]. \qquad (8)$$

Note that $\frac{d}{d\varphi}\Upsilon\boldsymbol{x}(\varphi) = 0$ for all $\varphi \in [0, 2\pi)$ when the residual $\boldsymbol{x} - \mathcal{P}\boldsymbol{x}$ is zero. Hence, the standard deviation of $\Upsilon\boldsymbol{x}$, denoted by $\text{std}(\Upsilon\boldsymbol{x})$, quantifies the amount of variation of the inconsistency. The crucial observations are the followings:

- High frequency mismatch $\text{std}(\Upsilon\boldsymbol{x}_H)$ (associated with abrupt change of $\Upsilon\boldsymbol{x}$ causes bright and dark streaking artifacts in the CT image.

- Low frequency mismatch $\text{std}(\Upsilon\boldsymbol{x}_L)$ (associated with a smoothly varying part of $\Upsilon\boldsymbol{x}$ with respect to $\varphi$) generates shading artifacts in the CT image.

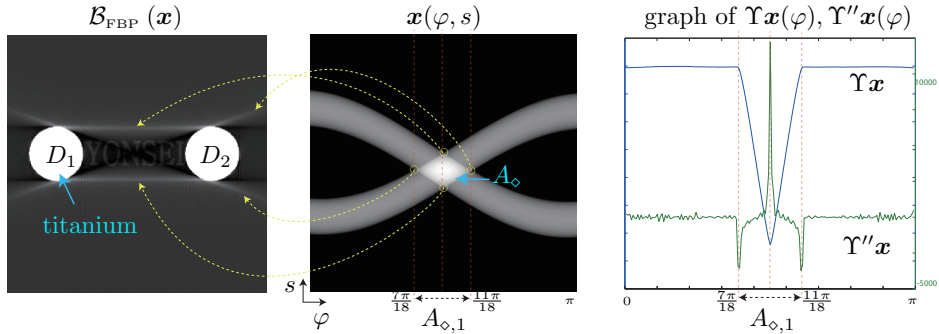

FIG. 4. Left figure shows the CT image $\mu_{\text{CT}}$ for numerical phantom containing two disk-shaped metallic objects (titanium) occupying the region $D_1$ and $D_2$. The attenuation coefficient of background image (marked by "YONSEI") is assumed to be the constant with energy $E$. Middle and right figures show projection data $\boldsymbol{x}$ and graph of $\Upsilon\boldsymbol{x}(\varphi)$ and $\frac{d^2}{d\varphi^2}\Upsilon\boldsymbol{x}(\varphi)$ of the phantom in the left figure, respectively. Here, CT image is simulated at energy range between 0-150 keV (C=-250 HU/W=2500 HU).

Fig. 4 illustrates the behavior of $\Upsilon\boldsymbol{x}(\varphi)$ and $\frac{d^2}{d\varphi^2}\Upsilon\boldsymbol{x}(\varphi)$ using a numerical phantom containing two disk-shaped metallic objects (titanium) occupying the region $D_1$ and $D_2$. In this simulation, we only consider the beam hardening artifacts for simplicity. The $\boldsymbol{x}$ is consistent in all the regions except the diamond shaped area $A_\diamond := \{(\varphi, s) : \mathcal{R}\chi_{D_1}(\varphi, s)\mathcal{R}\chi_{D_2}(\varphi, s) \neq 0\}$, where $\chi_D$ is the characteristic function, that is, $\chi_D = 1$ on $D$ and zero otherwise. This local inconsistency of $\boldsymbol{x}$ on $A_\diamond$ corresponds to variation of $\boldsymbol{\Upsilon x}$ on the interval $A_{\diamond,1} := \{\varphi : (\varphi, s) \in A_\diamond\}$. In the case of a single disk-shaped metallic object, the corresponding projection data $\boldsymbol{x}$ lies in the range space of the Radon transform[16]. However, $\mathcal{B}_{\text{FBP}}(\boldsymbol{x})$ contains cupping artifact inside the disk. Note that cupping artefacts only disturb the image in the regions of problematic objects, whereas streaking artefacts corrupt the tomographic image outside the problematic region[15].

From the inherent nature of the pseudo inverse of the Radon transform $\mathcal{R}$, the local inconsistency of $\boldsymbol{x}$ on $A_\diamond$ generates severe global artifacts in $\mathcal{B}_{\text{FBP}}(\boldsymbol{x})$, which appear as streaking and shading artifacts. Denoting the back-projection by $\mathcal{R}^*$ (the dual of $\mathcal{R}$), its mathematical structure can be explained as follows:

$$\mathcal{R}^*\mathcal{R}(\underset{\mathcal{B}_{\text{FBP}}(\boldsymbol{x})}{\boxed{\text{img}}}) = \mathcal{R}^*(\underset{\boldsymbol{x}}{\boxed{\text{img}}}) + \underbrace{\mathcal{R}^*(\boxed{\text{img}})}_{=0}$$

$$= \mathcal{R}^*(\underbrace{\boxed{\text{img}} - \boxed{\text{img}}}_{\text{corrected sinogram}}) + \mathcal{R}^*(\underbrace{\boxed{\text{img}} + \boxed{\text{img}}}_{\text{artifact sources}})$$

$$= \mathcal{R}^*\mathcal{R}(\boxed{\text{img}}) + \mathcal{R}^*\mathcal{R}(\boxed{\text{img}})$$

As illustrated in Fig. 4, the abrupt changes in $\boldsymbol{\Upsilon x}(\varphi)$ between projection views occur at $\varphi = \pi/2$ and $\varphi \approx 7\pi/18, 11\pi/18$. These abrupt discrepancies are mapped to the bright and dark streaking artifacts between the boundaries of two metallic objects in the $\mathcal{B}_{\text{FBP}}(\boldsymbol{x})$. Moreover, the gradually varying inconsistency in the $A_{\diamond,1} = [7\pi/18, 11\pi/18]$ interval causes shading artifacts between or near the two metallic objects.

## III. RESULTS

This section demonstrates the feasibility of the proposed method through realistic computer simulations. In our simulations, patient-type specific learning model was adopted, owing to difficulty in handling all possible metal geometries for training. This section focuses

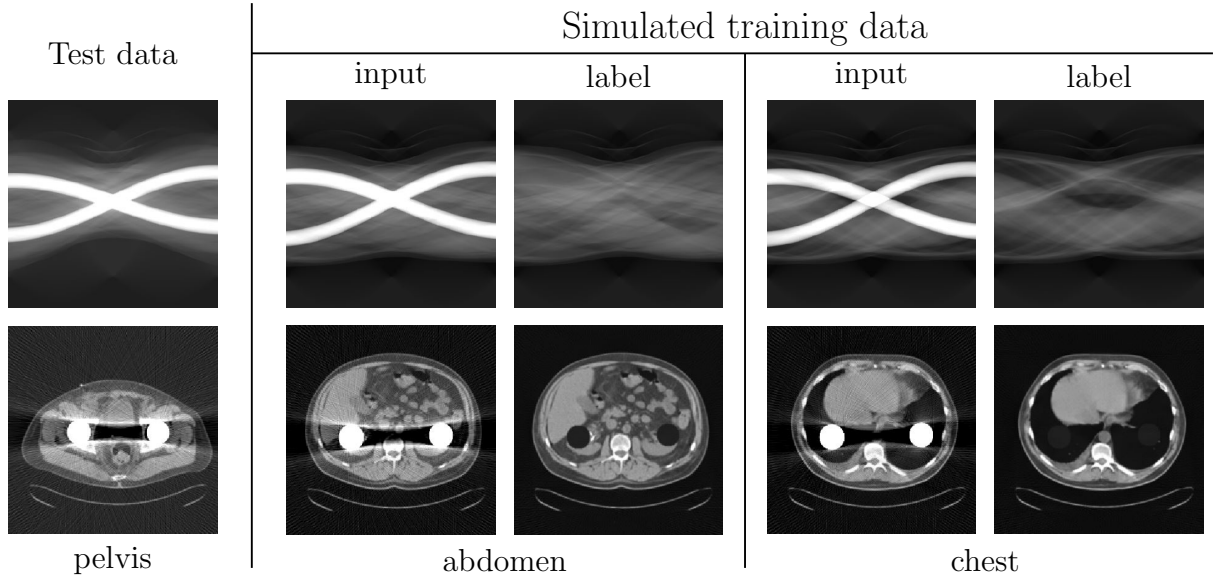

pelvis | abdomen | chest

FIG. 5. The anatomical areas in training and test data are different. The test data comes from a CT image of the pelvis, whereas the training sinograms are made from the CT images of abdomen and chest. We intentionally use these unrealistic training data to demonstrate that the proposed method effectively remove artifacts without deteriorating underlying morphological information.

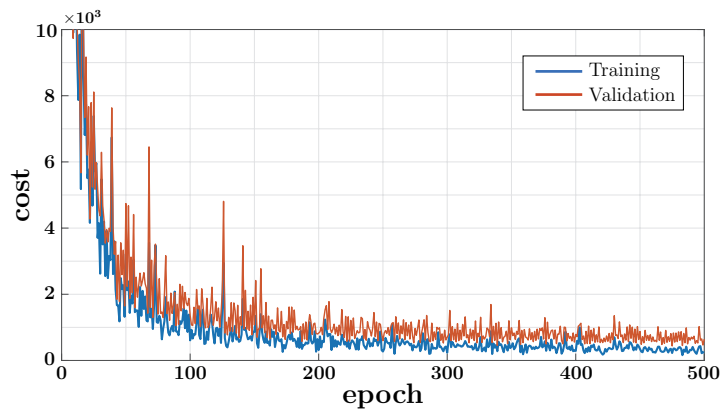

FIG. 6. Convergence plot of the cost function (6) for training and validation set with respect to training epochs.

on learning beam hardening artifacts caused by hip prosthesis. We inserted two simulated hip prostheses (made of iron) into CT images. Projection data $x$ containing Poisson noise was generated using the attenuation coefficient given in[9] and energy spectrum with 2.5 mm Al filtration at tube voltage of 150 kVp[19]. Here, we assumed that $\mu_E$ for pelvic CT images as a tissue. Other causes of metal artifacts, such as scattering and nonlinear partial volume effects were not considered.

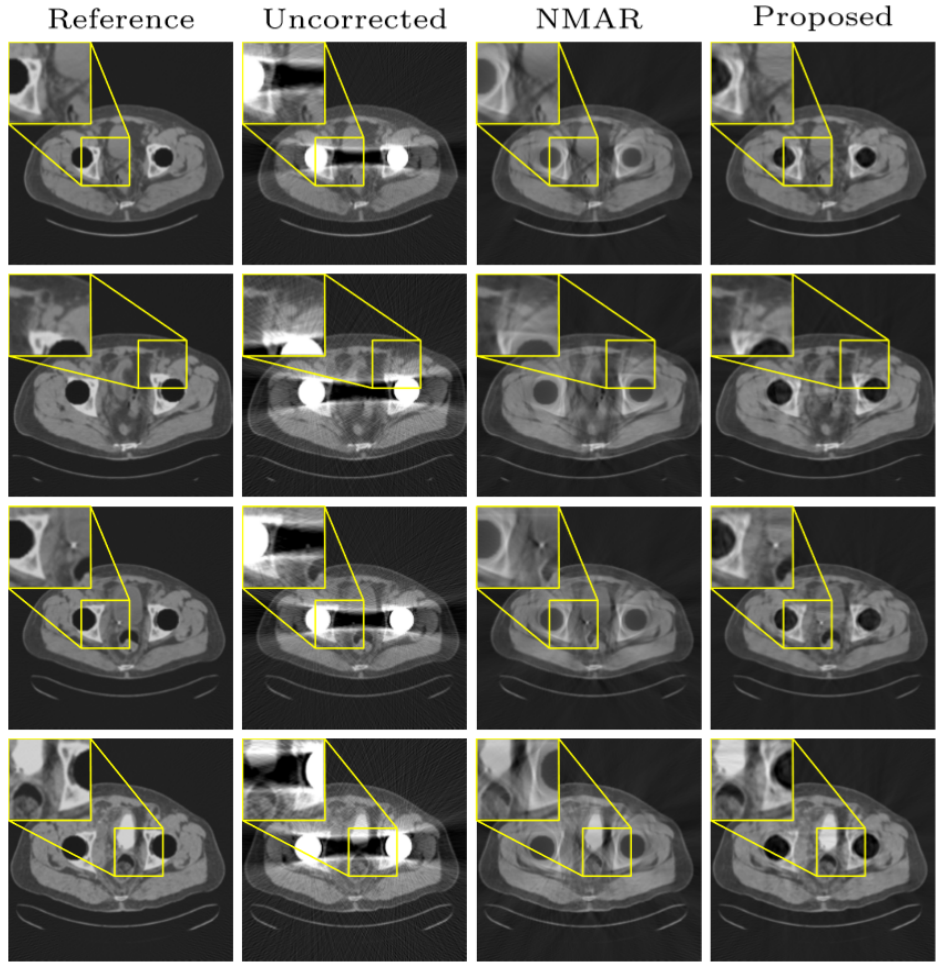

FIG. 7. Reconstruction results for pelvic CT images containing two simulated hip prostheses (made of iron). First and second columns show the reference and uncorrected image, respectively. The third and fourth columns represent the results with NMAR and the proposed method, respectively. First, second, third and fourth rows show results of variations of metal geometries with different tissue backgrounds. (C=-250 HU/W=2500 HU.)

For training, we intentionally did not use CT images of pelvis (test data's anatomical area), in order to demonstrate that the proposed method effectively extracts the beam-hardening factors without affecting artifact-free data. The input data is the projection data from CT images of chest and abdomen in the presence of hip prosthesis and the output is the corresponding forward projection data in the absence of hip prosthesis (See Fig. 5). To correct the main beam-hardening source along the metal trace, the method needs to apply exclusively on the metal trace. The metal trace is obtained by the forward projection of the metal region segmented from the uncorrected CT image.

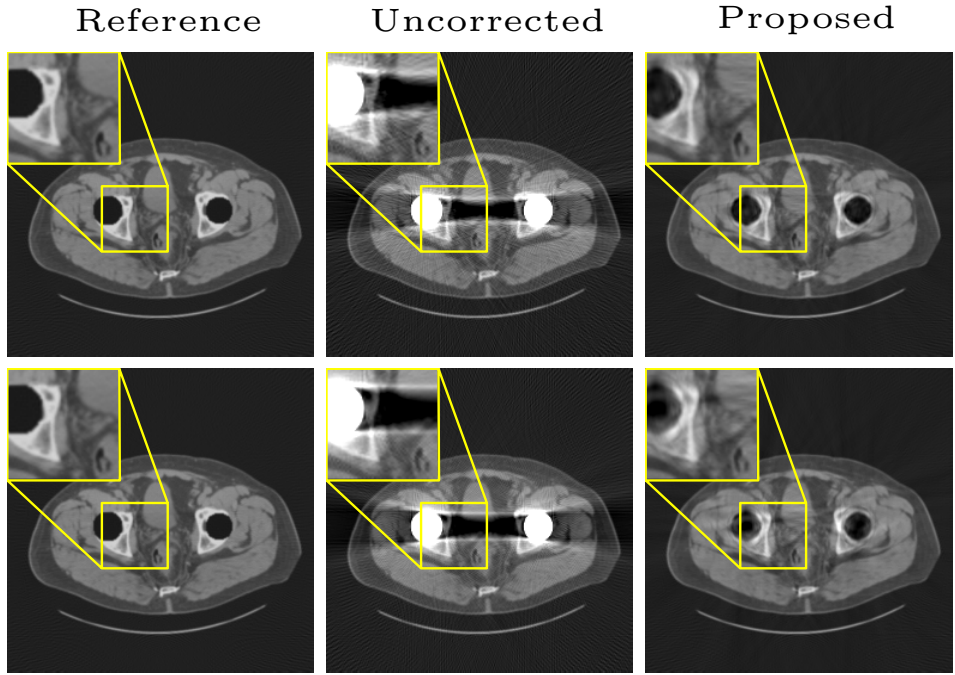

FIG. 8. Results of applying the proposed method to the projection data acquired in different environments from those used for training: the data generated by using the energy spectrum with 1.5 mm Al filtration at tube voltage of 150 kVp (first row) and data that includes the scattered photons (second row).

To train the network, we generated 3780 CT sinogram data with a size of 368 × 180 for CT images by changing position and size of the two metallic objects (iron). 3700 sinogram dataset were used for training and 80 sinogram dataset were used for evaluation. The cost function (6) was minimized using the RMSPropOptimizer[21] with a learning rate of 0.001, weight decay of 0.9, and mini-batch size of 32 at each epoch. Fig. 6 shows the convergence plot for the cost function (6) with respect to training epochs. We used a trained network with 500 training epochs. Training was implemented by Tensorflow[7] on a CPU (Intel(R) Core(TM) i7-6850K, 3.60GHz) and four GPU (NVIDIA GTX-1080, 8GB) system. The network required approximately 16 hours for training.

Fig. 7 shows the comparison between the proposed method and normalized metal artifact reduction (NMAR) method[13] for the simulated bi-lateral hip prostheses. As shown in the second column of Fig. 7, the uncorrected images suffer from severe beam hardening artifacts, which appear as streaking and shading artifacts between or near the two metallic objects. In addition, bright streaking artifacts occur in direction of metallic materials. This is because of noise due to insufficient photon reaching the detector after passing though the metallic objects[1]. The proposed and NMAR methods effectively reduce the streaking and shading artifacts arising from metallic objects. However, as shown in the third column of Fig. 7, the NMAR method does not recover the bone or tissue structures near the metallic object, since NMAR fills the corrupted metal trace from its surrounding information. Compared with NMAR method, the proposed method considerably recovers the morphological structure near the metallic objects (see fourth column of Fig. 7). This is because the proposed method tries to repair inconsistent sinograms by removing the primary metal-induced beam-hardening factors along the metal trace in the sinogram.

TABLE I. NRMSD of the reconstructed images with the NMAR method and the proposed method

|  | Phantom | Uncorrected | NMAR | Proposed |
|---|---|---|---|---|
| NRMSD (%) | Pelvis 1 | 38.3734 | 22.8523 | 8.1213 |
|  | Pelvis 2 | 47.8539 | 18.8835 | 10.0256 |
|  | Pelvis 3 | 47.4931 | 23.9704 | 8.7837 |
|  | Pelvis 4 | 46.8481 | 20.8512 | 9.3886 |

In order to estimate the quantitative errors of the uncorrected and corrected images, the normalized root mean square difference (NRMSD)[11] (with the label image) were computed on the outside of metals. The NRMSD (%) for each result is listed in Table I. In terms of NRMSD, overall errors of NMAR and the proposed methods are significantly reduced compared with that of uncorrected image. Compared with NMAR, the proposed method achieves a lower NRMSD.

Fig. 8 shows the results of applying the proposed method to the projection data acquired in different acquisition environments from those used for training: the data generated by using the energy spectrum with 1.5 mm Al filtration at tube voltage of 150 kVp (first row) and data that includes the scattered photons (second row). Specifically, the projection data

$x$ considering scatter is generated by

$$\tilde{\boldsymbol{x}}(\varphi,s) = -\ln\left(\int \eta(E)\exp\big\{-[\mathcal{R}(\mu_E)](\varphi,s)\big\} + \kappa_E(\varphi,s)dE\right), \qquad (9)$$

where $\kappa_E$ denotes the contribution of scatters. In this simulation, we assumed that $\kappa_E$ is a constant with respect to $(\varphi,s)$[5,12]. As shown in third column of Fig. 8, the proposed method reduces the metal induced artifacts while preserving the information near the metallic objects even for both data with slightly different energy spectrum and data with scatters. However, some streaking artifacts are still visible due to the inaccurate data along the metal trace.

## IV. DISCUSSION AND CONCLUSION

This paper proposed a deep learning method of sinogram correction for beam hardening reduction in CT for the first time. Conventional methods for beam hardening reduction are based on regularizations, and have the fundamental drawback of being not easily able to use manifold CT images, while a deep learning approach has the potential to do so. The fundamental challenge is to collect enough labeled training data, a necessary step for a deep learning process. We circumvent this issue by using simulated training data as the basis for learning sinogram repair. We use the fact that the geometry of a metal trace region in a sinogram is non-linearly related to the beam-hardening feature. We use a U-net to learn the map from the geometry of the metal trace region to the corresponding beam-hardening factor. Results demonstrate that the proposed method can effectively remove metal-induced inconsistencies.

Let us briefly discuss conventional regularization techniques such as compressed sensing using $\ell_1$-regularization. These approaches can effectively prevent inadmissible solutions, but they discard detailed information (e.g., they remove small anomalies containing clinically useful information for diagnosis), and hence they have limited applicability to computed medical imaging. Deep learning has the potential to learn non-linear regression for various artifact sources, because it effectively uses complicated prior knowledge of CT images and artifacts.

There is much room for improvement, and further research is necessary to deal with various metal and bone-related artifacts. This paper does not consider metal artifacts that arise from scattering, nonlinear partial volume effects or electric noise. The performance of the learning-based sinogram correction method could be improved by enhancing the forward model, which accurately represents various realistic artifacts. Our future research will include further investigation into learning approaches for sinogram correction and clinical research with patients.

ACKNOWLEDGEMENTS

This work was supported by Samsung Science & Technology Foundation (No. SSTF-BA1402-01). The first author was partially supported by the National Research Foundation of Korea(NRF) grant funded by the Korea government(Ministry of Science, ICT & Future Planning) (No. NRF-2016R1C1B2008098) and the National Institute for Mathematical Sciences (NIMS) grant funded by the Korean government (No. A21300000).

---

* Corresponding author: J.K. Seo (email: seoj@yonsei.ac.kr).